\documentclass[aps,prl,reprint,superscriptaddress]{revtex4-1}

\usepackage{graphicx}% Include figure files
\usepackage{dcolumn}% Align table columns on decimal point
\usepackage{bm}% bold math
\usepackage{siunitx}

\begin{document}

%\preprint{}

\title{Spin Hall Magnetoresistance in a Canted Ferrimagnet}
\author{Kathrin Ganzhorn}
\affiliation{Walther-Mei\ss ner-Institut, Bayerische Akademie der Wissenschaften, 85748 Garching, Germany}
\affiliation{Physik-Department, Technische Universit\"{a}t M\"{u}nchen, 85748 Garching, Germany}
\author{Joseph Barker}
\affiliation{Institute for Materials Research, Tohoku University, Sendai, Miyagi 980-8577, Japan}
\author{Richard Schlitz}
\author{Matthias Althammer}
\affiliation{Walther-Mei\ss ner-Institut, Bayerische Akademie der Wissenschaften, 85748 Garching, Germany}
\affiliation{Physik-Department, Technische Universit\"{a}t M\"{u}nchen, 85748 Garching, Germany}
\author{Stephan Gepr\"{a}gs}
\affiliation{Walther-Mei\ss ner-Institut, Bayerische Akademie der Wissenschaften, 85748 Garching, Germany}
\author{Hans Huebl}
\affiliation{Walther-Mei\ss ner-Institut, Bayerische Akademie der Wissenschaften, 85748 Garching, Germany}
\affiliation{Physik-Department, Technische Universit\"{a}t M\"{u}nchen, 85748 Garching, Germany}
\affiliation{Nanosystems Initiative Munich (NIM), Schellingstra\ss e 4, 80799 M\"{u}nchen, Germany}
\author{Benjamin A. Piot}
\affiliation{Laboratoire National des Champs Magn\'{e}tiques Intenses, LNCMI-CNRS-UGA-UPS-INSA-EMFL, F-38042 Grenoble, France}
\author{Rudolf Gross}
\affiliation{Walther-Mei\ss ner-Institut, Bayerische Akademie der Wissenschaften, 85748 Garching, Germany}
\affiliation{Physik-Department, Technische Universit\"{a}t M\"{u}nchen, 85748 Garching, Germany}
\affiliation{Nanosystems Initiative Munich (NIM), Schellingstra\ss e 4, 80799 M\"{u}nchen, Germany}
\author{Gerrit E. W. Bauer}
\affiliation{Institute for Materials Research, Tohoku University, Sendai, Miyagi 980-8577, Japan}
\affiliation{Kavli Institute of NanoScience, Delft University of Technology, 2628 CJ Delft, The Netherlands}
\affiliation{WPI Advanced Institute for Materials Research, Tohoku University, Sendai 980-8577, Japan}
\author{Sebastian T. B. Goennenwein}
\email{goennenwein@wmi.badw.de}
\affiliation{Walther-Mei\ss ner-Institut, Bayerische Akademie der Wissenschaften, 85748 Garching, Germany}
\affiliation{Physik-Department, Technische Universit\"{a}t M\"{u}nchen, 85748 Garching, Germany}
\affiliation{Nanosystems Initiative Munich (NIM), Schellingstra\ss e 4, 80799 M\"{u}nchen, Germany}

%\thanks{}


\date{\today}

\begin{abstract}
We study the spin Hall magnetoresistance effect in ferrimagnet/normal metal bilayers, comparing the response in collinear and canted magnetic phases. In the collinear magnetic phase, in which the sublattice magnetic moments are all aligned along the same axis, we observe the conventional spin Hall magnetoresistance. In contrast, in the canted phase, the magnetoresistance changes sign. Using atomistic spin model calculations of the magnetic configuration, we show that the electric transport for the different magnetic phases can be rationalized considering the individual sublattice moment orientations. This enables a magneto-transport based investigation of non-collinear magnetic textures.
\end{abstract}

%\keywords{}

\maketitle

The magnetic properties of ferromagnets are often modeled in terms of a simple macrospin with magnetization vector $\mathbf{M}$. In this picture, one tacitly assumes  that all individual atomic magnetic moments $\bm{\mu}$ are aligned in one direction, such that the magnetization is $\mathbf{M}=n \bm{\mu}$ with the moment number density $n$. However, many magnets exhibit a much richer magnetic structure, with canted, spiral, frustrated or even topological \cite{Dionne:Magnetic:Oxides:Book:2009,Muehlbauer:Skyrmions:Science:2009} phases appearing in addition to collinear magnetic order. Unravelling these experimentally typically requires sophisticated methods, e.g., spin polarized neutron scattering, x-ray magnetic circular dichroism, or Lorentz transmission electron microscopy. A pathway for the electrical detection of magnetic properties is provided by spin torques arising at a magnet/metal interface \cite{Kiselev:current-induced-M-reversal:Nature2004,spin-torque:Miron:NatMat:2010,Emori:Beach:STT:NatureMater:2013}. These torques govern fundamental spintronic phenomena such as spin pumping \cite{Urban:Woltersdorf:spin-currents:spin-pumping:PRL2001,spin-pumping:Tserkovnyak:PRL:2002,spin-pumping:Mosendz:PRL:2010,spin-pumping:Czeschka:scaling:PRL:2011,Takei:AFM:spincurrent:PRB:2015}, spin Seebeck effect \cite{longitudinal-spin-Seebeck:Uchida:APL:2010:172505,Weiler:local:SSE:PRL:2012,SSE:GdIG:Gepraegs:NatureCom:2016}, as well as spin Hall magnetoresistance \cite{NakayamaSMR,AltiPRB,ChenSMR,SMR:Hahn:PRB:2013,SMR:Vlietstra:PRB:2013}, and even enable an electrical control of the magnetization in magnetic nanostructures \cite{Kiselev:current-induced-M-reversal:Nature2004,spin-torque:Miron:NatMat:2010,Emori:Beach:STT:NatureMater:2013}. However, while the spin torque effect -- or more precisely the transfer of spin angular momentum across the magnet/metal interface -- has been extensively discussed for a macrospin $\mathbf{M}$ \cite{Slonczewski:1996,Berger1996}, the action of spin torques on non-collinear magnetic phases is only poorly understood.

In this Letter, we show that in the ferrimagnet gadolinium iron garnet ($\mathrm{Gd}_{3}\mathrm{Fe}_{5}\mathrm{O}_{12}$, GdIG), the spin Hall magnetoresistance (SMR) can be used to resolve the orientation of the individual atomic magnetic moments $\bm{\mu}_{X}$ residing on the different magnetic sublattices. We thereby prove that the SMR is not just governed by the net moment $\bm{\mu}_{\mathrm{net}}=\sum\bm{\mu}_{X}$ (viz.~the corresponding macrospin magnetization $\mathbf{M}_{\mathrm{net}}$) aligned along the externally applied magnetic field. This is reflected most conspicuously by the SMR sign inversion observed for canted sublattice moments. The interpretation of our experiments is corroborated by atomistic spin simulations \cite{SSE:GdIG:Gepraegs:NatureCom:2016}, suggesting that the Fe sublattice moments dominate the SMR response.

\begin{figure}
 \includegraphics[width=\columnwidth]{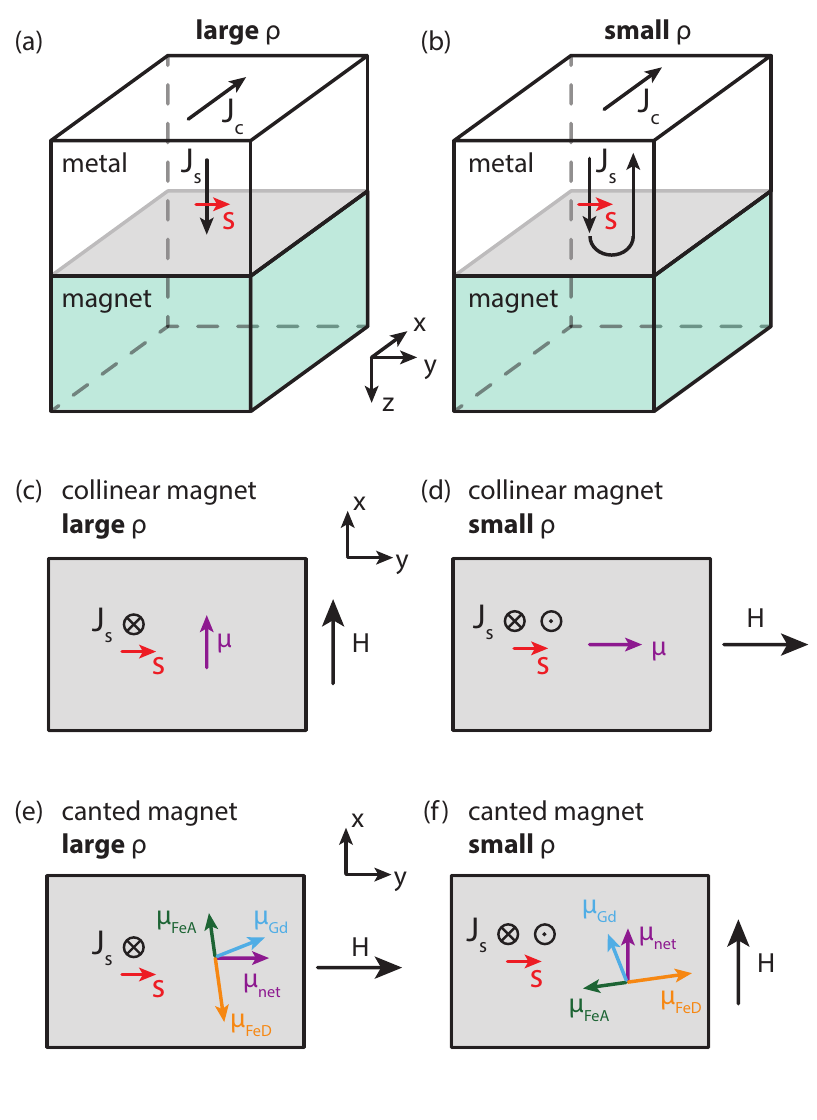}
 \caption{Spin Hall magnetoresistance (SMR) response of a magnetic insulator/metal bilayer. (a) When the spin current $\mathbf{J}_{\mathrm{s}}$ in the metal is absorbed by the magnet, the resistivity $\rho$ of the metal is large. (b) When $\mathbf{J}_{\mathrm{s}}$ is reflected at the interface, $\rho$ is small owing to the inverse spin Hall effect. (c),(d) In a collinear magnet, the spin transfer across the interface and thus $\rho$ is largest for $\bm{\mu}\perp \mathbf{s}$ (panel (c)), while spin transfer and $\rho$ are minimal for $\bm{\mu}||\mathbf{s}$ (panel (d)). (e),(f) In a non-collinear magnet in which, e.g., the orientation of the $\bm{\mu}_{\mathrm{FeA}}$ moments dominate the spin transfer across the interface, large viz.~small $\rho$ arises for the corresponding orientations of $\bm{\mu}_{\mathrm{FeA}}$ with respect to $\mathbf{s}$. An externally applied magnetic field $\mathbf{H}$ (larger than the weak anisotropy but smaller than the inter-spin exchange fields) determines the orientation of $\bm{\mu}_{\mathrm{net}}$. Comparing panels (c)-(f), the $\mathbf{H}$ orientations for maximum viz. minimum $\rho$ in the canted viz.~collinear magnet are interchanged -- the SMR inverts sign.}
 \label{Fig:SMRexpectation}
\end{figure}
The SMR originates from spin current transport across the interface between an (insulating) magnet and a metal with finite spin Hall angle. As sketched in Fig.\,\ref{Fig:SMRexpectation}(a), a charge current with density $\mathbf{J}_{\mathrm{c}}||\mathbf{x}$ induces a spin current density with direction $\mathbf{J}_{\mathrm{s}}||\mathbf{z}$ and polarization $\mathbf{s}||\mathbf{y}$ in the metal. Depending on whether $\mathbf{J}_{\mathrm{s}}$ is absorbed or reflected at the interface, the metal's resistivity $\rho$ is either increased (panel (a)) or not (panel (b)). In a collinear magnet, the amount of spin current at the interface can be modeled in terms of the magnetization direction $\mathbf{m}=\mathbf{M}/M=\bm{\mu}/\mu$ relative to $\mathbf{s}$. As sketched in panel (c), $\bm{\mu}\perp \mathbf{s}$ corresponds to maximal spin transfer and thus large $\rho$, while $\bm{\mu}||\mathbf{s}$ yields minimal $\rho$ (panel (d)), which can be parameterized by \cite{NakayamaSMR,AltiPRB,ChenSMR,SMR:review:Chen:JPCM:2016}
\begin{equation}\label{eq:SMR-1SL}
  \rho=\rho_0+\rho_{1} (\mathbf{m}\cdot\mathbf{y})^2.
\end{equation}

For more complex magnets, the use of Eq.\,(\ref{eq:SMR-1SL}) with $\mathbf{m}=\bm{\mu}_{\mathrm{net}}/\mu_{\mathrm{net}}$ becomes questionable. The magnet/metal exchange coupling in the SMR theory is formulated in terms of the spin mixing conductance, which for magnetic insulators is dominated by the local moments directly at the interface \cite{Jia:2011gm}. We can then illustrate the effect of the magnetization texture on the electron transport for a non-collinear magnet, viz.~the ferrimagnetic insulator GdIG with three magnetic sublattices ($\mathrm{FeA}$, $\mathrm{FeD}$ and $\mathrm{Gd}$) in a canted configuration as sketched in Fig.\,\ref{Fig:SMRexpectation}(e). Here, none of the local moments $\bm{\mu}_{\mathrm{FeA}}$, $\bm{\mu}_{\mathrm{FeD}}$, and $\bm{\mu}_{\mathrm{Gd}}$ are parallel to $\bm{\mu}_{\mathrm{net}}$. Since the antiferromagnetic exchange coupling between the FeA and FeD moments is strong, we sketch them as antiparallel in the figure. Therefore, $\bm{\mu}_{\mathrm{net}}\parallel \mathbf{H}$ is the vector sum of the net iron moment $\bm{\mu}_{\mathrm{FeD}}+\bm{\mu}_{\mathrm{FeA}}$ and of $\bm{\mu}_{\mathrm{Gd}}$.

To model the SMR in canted magnets, we assert that the spin-mixing conductance and the SMR is determined by the orientation of the individual, local magnetic moments $\bm{\mu}_{X}$ at the interface. The SMR then reads
 \begin{equation}\label{eq:SMR-ferri}
  \rho=\rho_0+\sum_{X} \rho_{1,X} \, \langle(\mathbf{m}_{X}\cdot\mathbf{y})^2 \rangle
\end{equation}
where $\langle \cdots \rangle$ denotes the average over all moments of type $X$, and $\rho_{1,X}$ is the corresponding SMR resistivity modulation.
For magnets with a collinear magnetization configuration, in which all sublattice moments are aligned parallel or antiparallel to each other, Eq.\,(\ref{eq:SMR-ferri}) is equivalent to Eq.\,(\ref{eq:SMR-1SL}) with $\rho_{1}=\sum_{X} \rho_{1,X}$. In other words, the SMR response of a collinear ferrimagnet according to Eq.\,(\ref{eq:SMR-ferri}) looks exactly like the SMR of a simple, one-sublattice ferromagnet. In contrast, for magnets with non-collinear spin structure, the SMR response depends on the orientations of the different sublattice moments in a non-trivial way.

\begin{figure}
 \includegraphics[width=\columnwidth]{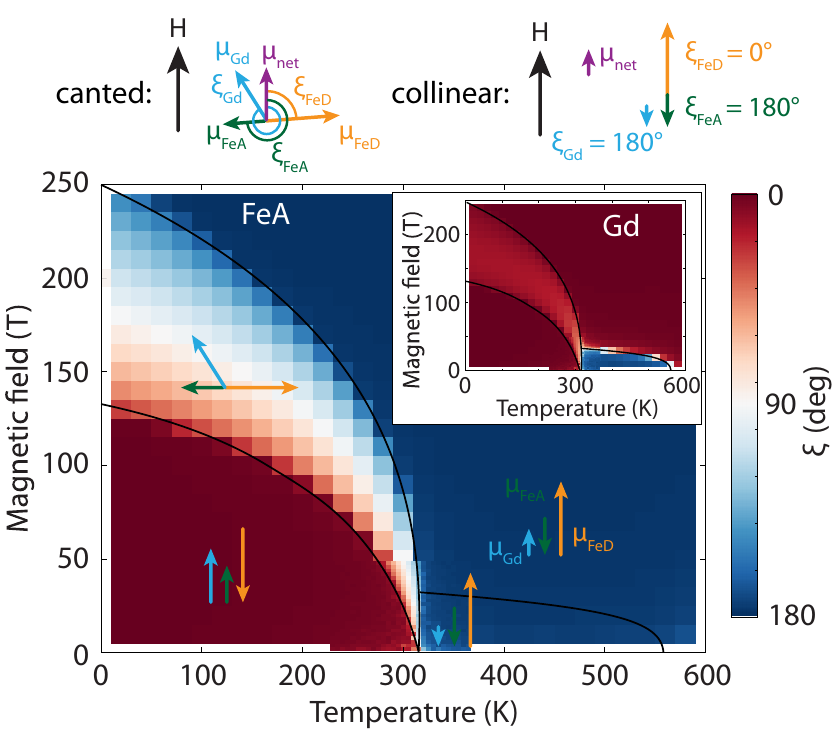}
 \caption{Magnetic phase diagram of GdIG calculated by atomistic spin simulations (see text). The main panel depicts the orientation of the $\mathrm{FeA}$ sublattice moment orientation $\xi_{\mathrm{FeA}}$ encoded in color, the inset shows $\xi_{\mathrm{Gd}}$ of the Gd sublattice moments. Due to the strong antiferromagnetic exchange coupling, the $\mathrm{FeD}$ sublattice moments are always antiparallel to the $\mathrm{FeA}$ ones. The black lines indicate the temperature dependence of the upper ($\mu_0 H_{c2}$) and lower ($\mu_0 H_{c1}$) critical fields which delimit the antiparallel, parallel and spin canting phase \cite{Bernasconi, Clark}. The orientation of the sublattice moments in each phase are represented by arrows.}
 \label{Fig:phase}
 \end{figure}

Most SMR experiments to date have been performed on bilayers made from yttrium iron garnet ($\mathrm{Y}_{3}\mathrm{Fe}_{5}\mathrm{O}_{12}$, YIG) as the insulating magnet and platinum (Pt) as the metal. The magnetic properties of YIG stem from two octahedrally coordinated Fe$^{3+}$ moments ($\mathrm{FeA}$) and three tetrahedrally coordinated Fe$^{3+}$ moments ($\mathrm{FeD}$) per formula unit. The $\mathrm{FeA}$ and $\mathrm{FeD}$ moments are strongly antiferromagnetically coupled. YIG therefore is a collinear ferrimagnet, warranting the use of Eq.\,(\ref{eq:SMR-1SL}). Only in magnetic fields in excess of $\mu_0 H_{c1}\approx 250\,\mathrm{T}$, a canted magnetic phase emerges, in which $\mathbf{H}$, $\bm{\mu}_{\mathrm{FeA}}$ and $\bm{\mu}_{\mathrm{FeD}}$ are no longer aligned along one common axis \cite{Chikazumi,Bernasconi, Clark,Dionne:Magnetic:Oxides:Book:2009}.

In contrast to YIG, the canted magnetic phase is readily accessible in compensated magnetic garnets such as GdIG, see Fig.\,\ref{Fig:phase}. Due to their exchange coupling to the $\mathrm{FeA}$ and $\mathrm{FeD}$ moments, the paramagnetic Gd moments acquire a finite sublattice magnetization \cite{Dionne:Magnetic:Oxides:Book:2009}. We model the GdIG magnetic structure using a classical Heisenberg Hamiltonian including all of the atoms in the unit cell (see Ref~\cite{SSE:GdIG:Gepraegs:NatureCom:2016} for details of the model). We use a Metropolis Monte Carlo algorithm with a combination of different moves to avoid trapping in metastable minima \cite{Hinzke:1999ud}, to calculate the equilibrium magnetic configuration as a function of applied field and temperature, disregarding the small crystalline anisotropy. The system size is $16 \times 16 \times 16$ unit cells (131072 spins) with periodic boundary conditions. In particular, we take the spin configuration at the surface to be  similar to that of the bulk. Figure \ref{Fig:phase} shows the (average) orientation $\xi_{\mathrm{FeA}}$ of the $\mathrm{FeA}$ sublattice moments with respect to the applied field direction in the main panel, as well as the orientation $\xi_{\mathrm{Gd}}$ of the Gd ones in the inset. Since the $\mathrm{FeA}$ and $\mathrm{FeD}$ moments are coupled via a strong antiferromagnetic exchange, $\xi_{\mathrm{FeD}}=\xi_{\mathrm{FeA}}+\SI{180}{\degree}$. As evident from Fig.\,\ref{Fig:phase}, the $\mathrm{FeA}$, $\mathrm{FeD}$ and Gd sublattices arrange in different configurations depending on temperature and external magnetic field. Moreover, a magnetic compensation point with $\mathbf{M}_{\mathrm{net}}=0$ for $H=0$ arises at the so-called compensation temperature $T_{\mathrm{comp}}\approx \SI{300}{K}$. A canted magnetic phase is easily accessible already for magnetic fields of a few Tesla in the vicinity of $T_{\mathrm{comp}}$. $T_{\mathrm{comp}}$ and the critical fields are reduced by alloying In and Y into GdIG, so that a large portion of the canted phase becomes accessible using standard magnet cryostats. SMR experiments in InYGdIG/Pt bilayers thus are an ideal testbed to check the validity of Eq.\,(\ref{eq:SMR-ferri}).

We here discuss experiments on two different garnet/Pt bilayers. The magnetic garnet layers were deposited onto single crystalline, [111]-oriented Yttrium Aluminum Garnet (Y$_3$Al$_5$O$_{12}$, YAG) substrates via pulsed laser deposition (PLD).
The Yttrium Iron Garnet (Y$_3$Fe$_5$O$_{12}$, YIG) film was grown using a substrate temperature of \SI{500}{\celsius}, an oxygen atmosphere of \SI{2.5e-2}{\milli\bar}, and an energy fluence of the KrF excimer laser of \SI[per-mode=symbol]{2.0}{\joule\per\square\centi\meter} at the target surface. The \SI{40}{\nano\meter} thick YIG film was covered \textit{in-situ} with $t=\SI{4}{\nano\meter}$ of Pt deposited via electron beam evaporation. The same growth parameters were used for the Indium and Yttrium doped Gadolinium Iron Garnet (Y$_1$Gd$_2$Fe$_4$In$_1$O$_{12}$, InYGdIG) film, which has a thickness of \SI{61.5}{\nano\meter} and was covered with $t=\SI{3.6}{\nano\meter}$ of Pt. The InYGdIG sample exhibits a magnetization compensation temperature $T_\mathrm{comp}=\SI{85}{\kelvin}$, such that magnetotransport experiments at temperatures well above and well below $T_\mathrm{comp}$ are possible in standard magnet cryostats. Both garnet/Pt bilayers were patterned into Hall bars with width $w=\SI{80}{\micro\meter}$ and length $l=\SI{600}{\micro\meter}$ using optical lithography and argon ion beam milling.

For magnetoresistance measurements at magnetic fields up to $\mu_0 H = \SI{7}{\tesla}$ the samples were mounted in the variable temperature insert of a superconducting magnet cryostat ($\SI{10}{\kelvin} \leq T \leq \SI{300}{\kelvin}$) at the Walther-Meissner-Institut (WMI). Additional measurements up to $\mu_0 H = \SI{29}{\tesla}$ were conducted using a resistive magnet setup with a variable temperature insert at the high-field magnet laboratory in Grenoble. In both setups, a constant current of $I=\SI{0.2}{\milli\ampere}$ was applied along the Hall bar using a Keithley 2400 sourcemeter. We carried out angle-dependent magnetoresistance measurements \cite{AltiPRB} by rotating the sample with respect to the external magnetic field of fixed magnitude $\mu_0 H \le \SI{29}{\tesla}$ applied in the sample plane, simultaneously recording the voltage drop $V$ along the direction of charge current as a function of the angle $\alpha_{H}$ between the current direction $\mathbf{J}_{\mathrm{c}}$ and the external magnetic field $\textbf{H}$ using a Keithley 2182 nanovoltmeter. Hereby, we used a current reversal method in order to cancel thermopower effects and reduce noise.

\begin{figure}%
\includegraphics{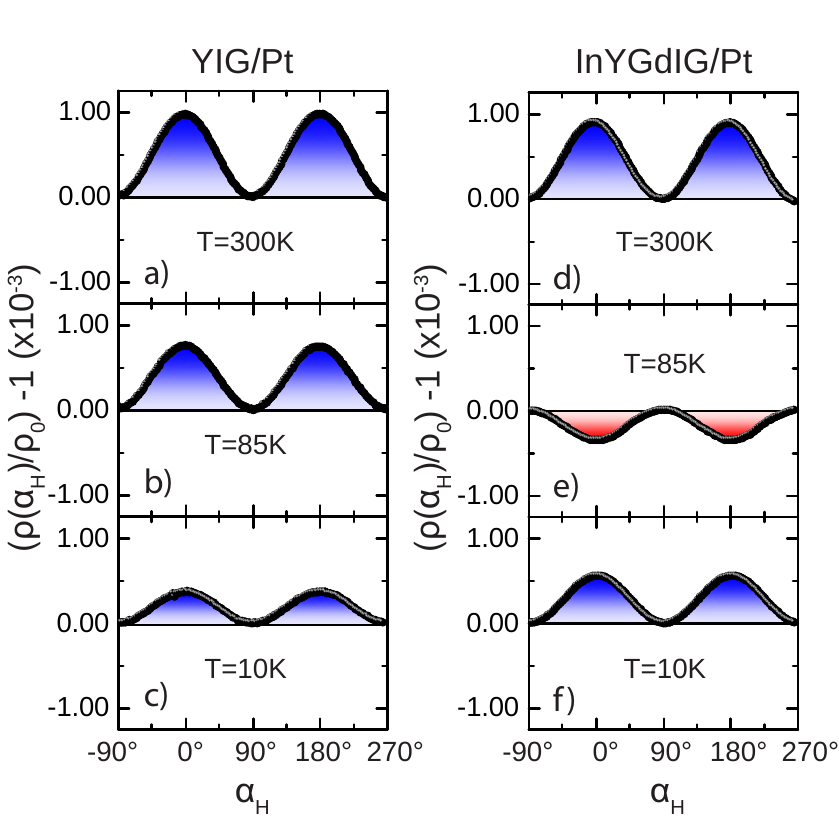}%
\caption{Measured evolution of the magnetoresistance in YIG/Pt (panels a-c) and InYGdIG/Pt (panels d-f). The data were recorded at $T=\SI{10}{\kelvin}, \SI{85}{\kelvin}$ and $\SI{300}{\kelvin}$ as a function of the angle $\alpha_{H}$ between the current direction $\textbf{J}_{\mathrm{c}}$ and the orientation of the external, in-plane magnetic field $\mu_0 H=\SI{7}{T}$, respectively. The SMR in InYGdIG/Pt inverts sign around the magnetization compensation temperature $T_\mathrm{comp}\approx \SI{85}{\kelvin}$ (panel e), but the extrema stay at the same $\alpha_{H}$ for all temperatures.}%
\label{Fig:admr}%
\end{figure}

Figure \ref{Fig:admr}(a-c) shows a typical set of magnetoresistance measurements for the YIG/Pt bilayer, taken at fixed temperatures of $T=\SI{10}{\kelvin}, \SI{85}{\kelvin}, \SI{300}{\kelvin}$ and an external magnetic field of \SI{7}{\tesla}. The magnetoresistance behavior is fully consistent with previous measurements \cite{AltiPRB,MeyerSMRAPL}.
Indeed, taking both the applied magnetic field $\mathbf{H}$ and $\mathbf{m}$ to reside in the magnet/metal interface plane (the $\mathbf{xy}$ plane in Fig.\,\ref{Fig:SMRexpectation}), Eq.\,(\ref{eq:SMR-1SL}) can be rewritten as $\rho(\alpha_{H})=\rho_0+\rho_{1} \sin^2 \alpha_{H}$, with $\rho_{1}<0$ \cite{SMR:review:Chen:JPCM:2016}.
The SMR amplitude
\begin{equation}
\frac{-\rho_{1}}{\rho\left(\alpha_{H}=\SI{0}{\degree}\right)} =\frac{\rho\left(\alpha_{H}=\SI{0}{\degree}\right)-\rho\left(\alpha_{H}=\SI{90}{\degree}\right)}{\rho\left(\alpha_{H}=\SI{0}{\degree}\right)}
\label{eq:delta rho}
\end{equation}
is positive at all temperatures, and decreases with decreasing temperature by about a factor of 2 as also reported in the literature \cite{MeyerSMRAPL}. A similar set of magnetoresistance measurements for the InYGdIG/Pt sample is depicted in Fig.~\ref{Fig:admr}(d-f), again for $T=\SI{10}{\kelvin}, \SI{85}{\kelvin}$ and $\SI{300}{\kelvin}$. The measurements at $T=\SI{10}{\kelvin}$ and $T=\SI{300}{\kelvin}$ (panels (d) and (f)) show the same positive SMR as for YIG/Pt. However, at $T=\SI{85}{\kelvin}\approx T_\mathrm{comp}$ (panel (e)), the SMR has negative sign, and comparatively small amplitude. This is surprising and cannot be accounted for by the standard SMR theory as written in Eq.\,(\ref{eq:SMR-1SL}) \cite{ChenSMR,SMR:review:Chen:JPCM:2016}.

To substantiate the SMR sign change,
\begin{figure}
\center
 \includegraphics[width=\columnwidth]{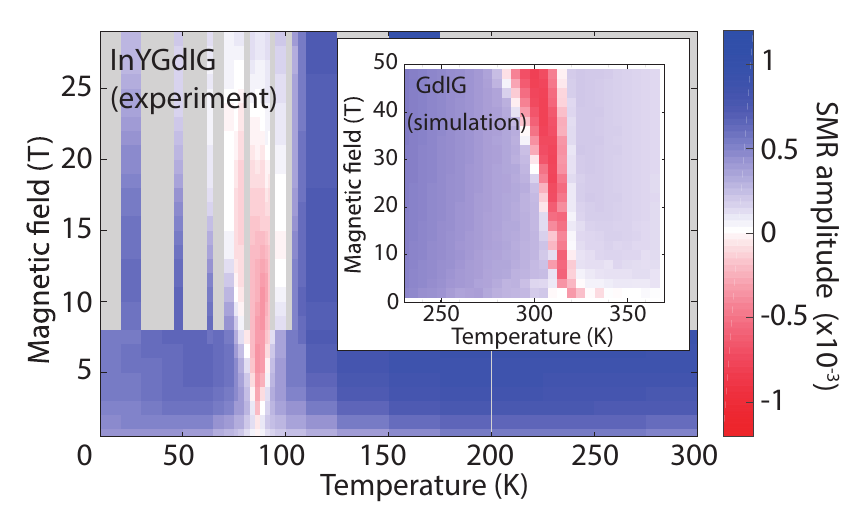}%
 \caption{SMR amplitude Eq.\,(\ref{eq:delta rho}) as a function of temperature and magnetic field, as measured for InYGdIG (main figure), and calculated for GdIG (inset) only taking the iron moments into account. In the blue regions the SMR is positive, i.e. has the same sign and $\alpha_{H}$-dependence as for a single-sublattice ferromagnet (cf. Fig.\,\ref{Fig:SMRexpectation}(a)). The red regions indicate negative SMR (as in Fig.\,\ref{Fig:admr}(e)). No data has been taken in the regions shaded in grey.}
 \label{Fig:SMRvsTandB}
 \end{figure}
we studied the evolution of the SMR amplitude (Eq.\,(\ref{eq:delta rho})) with magnetic field strength and temperature in the YIG/Pt and InYGdIG/Pt samples. In YIG/Pt, the SMR amplitude monotonically increases with $T$, as reported previously \cite{MeyerSMRAPL}. In InYGdIG/Pt, the behaviour is much richer. Figure \ref{Fig:SMRvsTandB} shows corresponding data obtained for $\mu_0 H\le\SI{7}{T}$ in the superconducting magnet cryostat at WMI, as well as $\mu_0 H\le\SI{29}{T}$ at the high field magnet laboratory in Grenoble, in a false color plot. The SMR sign change in InYGdIG/Pt is clearly evident as a red pocket around $T_\mathrm{comp} \mathrm{(InYGdIG)}=\SI{85}{\kelvin}$.

We may conclude with confidence that the macrospin picture of the SMR breaks down for non-collinear magnets. Since the spin current transport across the magnetic insulator/normal metal interface relevant for SMR corresponds to an additional (transverse) dissipation channel for charge transport, $\rho_{1} <0$ cannot change sign with temperature \cite{ChenSMR,AltiPRB}. The large external magnetic fields well exceed the demagnetizing or anisotropy fields, such that the orientation $\alpha_{H}$ of $\mu_0 H$ is identical to the orientation of $\bm{\mu}_{\mathrm{net}}$ viz.\,$\mathbf{M}_{\mathrm{net}}$. Thus, if $\mathbf{m}=\bm{\mu}_{\mathrm{net}}/\mu_{\mathrm{net}}$ indeed would govern the SMR in the spirit of Eq.\,(\ref{eq:SMR-1SL}), the SMR amplitude should be positive for all temperatures and magnetic fields. The InYGdIG/Pt sample clearly violates this conjecture, showing that the SMR is a powerful method to characterize complex spin textures. The small SMR modulation observed in CoCr$_2$O$_4$ can thus indeed be evidence for different spin spiral phases \cite{SMR:CCO:Aqeel:PRB2015}.

In the following, we show that the SMR response summarized in Fig.\,\ref{Fig:SMRvsTandB} can be straightforwardly understood assuming that the magnetic sublattice moments contribute independently to the SMR, as expressed in Eq.\,(\ref{eq:SMR-ferri}). Since the exchange parameters of InYGdIG are not well known, we compare the experimental SMR data from InYGdIG/Pt with the SMR calculated for GdIG/Pt (Fig.\,\ref{Fig:SMRvsTandB}). While the compensation temperatures of GdIG and InYGdIG are different, the spin correlations and thus the canted phases should be qualitatively similar. Indeed, the SMR response calculated from Eqs.\,(\ref{eq:SMR-ferri}) and (\ref{eq:delta rho}) using the sublattice moment orientations $\xi_{X}$ from Fig.\,\ref{Fig:phase} reproduces all the salient features observed in experiment. Interestingly, a reasonable agreement between model and experiment is obtained already upon taking into account only the iron moments, as shown in the inset of Fig.\,\ref{Fig:SMRvsTandB}. The Gd moments play a minor role for the SMR in GdIG, owing to a large spread in their directions arising from thermal fluctuations. Assuming that the iron sublattice moment orientations govern the SMR, we can understand its sign reversal in the canted phase from Fig.\,\ref{Fig:SMRexpectation}. While in the collinear phase the iron (and also the Gd) moments are aligned along the $\mathbf{H}$ axis, they rotate away from $\mathbf{H}$ in the canted phase. As indicated in the figure, this reorientation of the iron magnetic moments relative to the applied magnetic field causes the inversion of the SMR. We note that a given $\bm{\mu}_{\mathrm{net}}$ can result from different $\bm{\mu}_{X}$ textures. While different sublattice moment configurations are naturally included in the atomistic modelling used here, their impact on the SMR warrants a more detailed study in the future.

In summary, we observe a sign inversion of the SMR in compensated ferrimagnet/Pt bilayers around $T_\mathrm{comp}$. We attribute this behaviour to the non-collinear reorientation of the sublattice moments in the spin canting phase. We show that the experimental data can be understood assuming that the magnetic moments in the different magnetic sublattices contribute independently to the SMR. Our results demonstrate that simple transport experiments can identify non-collinear magnetic phases in highly resistive magnets contacted by heavy metals. The SMR thus might prove useful also for the investigation of topological spin textures, e.g., skyrmions, in thin films and nanostructures.

We thank Francesco Della Coletta, Sibylle Meyer, and Sascha Fr\"{o}lich for sample fabrication and gratefully acknowledge financial support via DFG Priority Programme 1538 “Spin-Caloric Transport” (GO 944/4, BA 2954/2), EU FP7 ICT Grant No. 612759 InSpin, and Grant-in-Aid for Scientific Research (Grant Nos. 25247056, 25220910, 26103006). J.B. acknowledges support from the Graduate Program in Spintronics, Tohoku University.

%\bibliography{bibliography-SMR-canting}

\begin{thebibliography}{28}%
\makeatletter
\providecommand \@ifxundefined [1]{%
 \@ifx{#1\undefined}
}%
\providecommand \@ifnum [1]{%
 \ifnum #1\expandafter \@firstoftwo
 \else \expandafter \@secondoftwo
 \fi
}%
\providecommand \@ifx [1]{%
 \ifx #1\expandafter \@firstoftwo
 \else \expandafter \@secondoftwo
 \fi
}%
\providecommand \natexlab [1]{#1}%
\providecommand \enquote  [1]{``#1''}%
\providecommand \bibnamefont  [1]{#1}%
\providecommand \bibfnamefont [1]{#1}%
\providecommand \citenamefont [1]{#1}%
\providecommand \href@noop [0]{\@secondoftwo}%
\providecommand \href [0]{\begingroup \@sanitize@url \@href}%
\providecommand \@href[1]{\@@startlink{#1}\@@href}%
\providecommand \@@href[1]{\endgroup#1\@@endlink}%
\providecommand \@sanitize@url [0]{\catcode `\\12\catcode `\$12\catcode
  `\&12\catcode `\#12\catcode `\^12\catcode `\_12\catcode `\%12\relax}%
\providecommand \@@startlink[1]{}%
\providecommand \@@endlink[0]{}%
\providecommand \url  [0]{\begingroup\@sanitize@url \@url }%
\providecommand \@url [1]{\endgroup\@href {#1}{\urlprefix }}%
\providecommand \urlprefix  [0]{URL }%
\providecommand \Eprint [0]{\href }%
\providecommand \doibase [0]{http://dx.doi.org/}%
\providecommand \selectlanguage [0]{\@gobble}%
\providecommand \bibinfo  [0]{\@secondoftwo}%
\providecommand \bibfield  [0]{\@secondoftwo}%
\providecommand \translation [1]{[#1]}%
\providecommand \BibitemOpen [0]{}%
\providecommand \bibitemStop [0]{}%
\providecommand \bibitemNoStop [0]{.\EOS\space}%
\providecommand \EOS [0]{\spacefactor3000\relax}%
\providecommand \BibitemShut  [1]{\csname bibitem#1\endcsname}%
\let\auto@bib@innerbib\@empty
%</preamble>
\bibitem [{\citenamefont {Dionne}(2009)}]{Dionne:Magnetic:Oxides:Book:2009}%
  \BibitemOpen
  \bibfield  {author} {\bibinfo {author} {\bibfnamefont {G.~F.}\ \bibnamefont
  {Dionne}},\ }\href@noop {} {\emph {\bibinfo {title} {Magnetic Oxides}}}\
  (\bibinfo  {publisher} {Springer},\ \bibinfo {address} {New York},\ \bibinfo
  {year} {2009})\BibitemShut {NoStop}%
\bibitem [{\citenamefont {M\"uhlbauer}\ \emph {et~al.}(2009)\citenamefont
  {M\"uhlbauer}, \citenamefont {Binz}, \citenamefont {Jonietz}, \citenamefont
  {Pfleiderer}, \citenamefont {Rosch}, \citenamefont {Neubauer}, \citenamefont
  {Georgii},\ and\ \citenamefont {B\"oni}}]{Muehlbauer:Skyrmions:Science:2009}%
  \BibitemOpen
  \bibfield  {author} {\bibinfo {author} {\bibfnamefont {S.}~\bibnamefont
  {M\"uhlbauer}}, \bibinfo {author} {\bibfnamefont {B.}~\bibnamefont {Binz}},
  \bibinfo {author} {\bibfnamefont {F.}~\bibnamefont {Jonietz}}, \bibinfo
  {author} {\bibfnamefont {C.}~\bibnamefont {Pfleiderer}}, \bibinfo {author}
  {\bibfnamefont {A.}~\bibnamefont {Rosch}}, \bibinfo {author} {\bibfnamefont
  {A.}~\bibnamefont {Neubauer}}, \bibinfo {author} {\bibfnamefont
  {R.}~\bibnamefont {Georgii}}, \ and\ \bibinfo {author} {\bibfnamefont
  {P.}~\bibnamefont {B\"oni}},\ }\href {\doibase 10.1126/science.1166767}
  {\bibfield  {journal} {\bibinfo  {journal} {Science}\ }\textbf {\bibinfo
  {volume} {323}},\ \bibinfo {pages} {915} (\bibinfo {year}
  {2009})}\BibitemShut {NoStop}%
\bibitem [{\citenamefont {Kiselev}\ \emph {et~al.}(2003)\citenamefont
  {Kiselev}, \citenamefont {Sankey}, \citenamefont {Krivorotov}, \citenamefont
  {Emley}, \citenamefont {Schoelkopf}, \citenamefont {Buhrman},\ and\
  \citenamefont {Ralph}}]{Kiselev:current-induced-M-reversal:Nature2004}%
  \BibitemOpen
  \bibfield  {author} {\bibinfo {author} {\bibfnamefont {S.~I.}\ \bibnamefont
  {Kiselev}}, \bibinfo {author} {\bibfnamefont {J.~C.}\ \bibnamefont {Sankey}},
  \bibinfo {author} {\bibfnamefont {I.~N.}\ \bibnamefont {Krivorotov}},
  \bibinfo {author} {\bibfnamefont {N.~C.}\ \bibnamefont {Emley}}, \bibinfo
  {author} {\bibfnamefont {R.~J.}\ \bibnamefont {Schoelkopf}}, \bibinfo
  {author} {\bibfnamefont {R.~A.}\ \bibnamefont {Buhrman}}, \ and\ \bibinfo
  {author} {\bibfnamefont {D.~C.}\ \bibnamefont {Ralph}},\ }\href@noop {}
  {\bibfield  {journal} {\bibinfo  {journal} {Nature}\ }\textbf {\bibinfo
  {volume} {425}},\ \bibinfo {pages} {380} (\bibinfo {year}
  {2003})}\BibitemShut {NoStop}%
\bibitem [{\citenamefont {Miron}\ \emph {et~al.}(2010)\citenamefont {Miron},
  \citenamefont {Gaudin}, \citenamefont {Auffret}, \citenamefont {Rodmacq},
  \citenamefont {Schuhl}, \citenamefont {Pizzini}, \citenamefont {Vogel},\ and\
  \citenamefont {Gambardella}}]{spin-torque:Miron:NatMat:2010}%
  \BibitemOpen
  \bibfield  {author} {\bibinfo {author} {\bibfnamefont {I.~M.}\ \bibnamefont
  {Miron}}, \bibinfo {author} {\bibfnamefont {G.}~\bibnamefont {Gaudin}},
  \bibinfo {author} {\bibfnamefont {S.}~\bibnamefont {Auffret}}, \bibinfo
  {author} {\bibfnamefont {B.}~\bibnamefont {Rodmacq}}, \bibinfo {author}
  {\bibfnamefont {A.}~\bibnamefont {Schuhl}}, \bibinfo {author} {\bibfnamefont
  {S.}~\bibnamefont {Pizzini}}, \bibinfo {author} {\bibfnamefont
  {J.}~\bibnamefont {Vogel}}, \ and\ \bibinfo {author} {\bibfnamefont
  {P.}~\bibnamefont {Gambardella}},\ }\href {\doibase 10.1038/nmat2613}
  {\bibfield  {journal} {\bibinfo  {journal} {Nat. Mater.}\ }\textbf {\bibinfo
  {volume} {9}},\ \bibinfo {pages} {230} (\bibinfo {year} {2010})}\BibitemShut
  {NoStop}%
\bibitem [{\citenamefont {Emori}\ \emph {et~al.}(2013)\citenamefont {Emori},
  \citenamefont {Bauer}, \citenamefont {Ahn}, \citenamefont {Martinez},\ and\
  \citenamefont {Beach}}]{Emori:Beach:STT:NatureMater:2013}%
  \BibitemOpen
  \bibfield  {author} {\bibinfo {author} {\bibfnamefont {S.}~\bibnamefont
  {Emori}}, \bibinfo {author} {\bibfnamefont {U.}~\bibnamefont {Bauer}},
  \bibinfo {author} {\bibfnamefont {S.-M.}\ \bibnamefont {Ahn}}, \bibinfo
  {author} {\bibfnamefont {E.}~\bibnamefont {Martinez}}, \ and\ \bibinfo
  {author} {\bibfnamefont {G.~S.~D.}\ \bibnamefont {Beach}},\ }\href {\doibase
  10.1038/nmat3675} {\bibfield  {journal} {\bibinfo  {journal} {Nature
  Materials}\ }\textbf {\bibinfo {volume} {12}},\ \bibinfo {pages} {611}
  (\bibinfo {year} {2013})}\BibitemShut {NoStop}%
\bibitem [{\citenamefont {Urban}\ \emph {et~al.}(2001)\citenamefont {Urban},
  \citenamefont {Woltersdorf},\ and\ \citenamefont
  {Heinrich}}]{Urban:Woltersdorf:spin-currents:spin-pumping:PRL2001}%
  \BibitemOpen
  \bibfield  {author} {\bibinfo {author} {\bibfnamefont {R.}~\bibnamefont
  {Urban}}, \bibinfo {author} {\bibfnamefont {G.}~\bibnamefont {Woltersdorf}},
  \ and\ \bibinfo {author} {\bibfnamefont {B.}~\bibnamefont {Heinrich}},\
  }\href {\doibase 10.1103/PhysRevLett.87.217204} {\bibfield  {journal}
  {\bibinfo  {journal} {Phys. Rev. Lett.}\ }\textbf {\bibinfo {volume} {87}},\
  \bibinfo {pages} {217204} (\bibinfo {year} {2001})}\BibitemShut {NoStop}%
\bibitem [{\citenamefont {Tserkovnyak}\ \emph {et~al.}(2002)\citenamefont
  {Tserkovnyak}, \citenamefont {Brataas},\ and\ \citenamefont
  {Bauer}}]{spin-pumping:Tserkovnyak:PRL:2002}%
  \BibitemOpen
  \bibfield  {author} {\bibinfo {author} {\bibfnamefont {Y.}~\bibnamefont
  {Tserkovnyak}}, \bibinfo {author} {\bibfnamefont {A.}~\bibnamefont
  {Brataas}}, \ and\ \bibinfo {author} {\bibfnamefont {G.~E.~W.}\ \bibnamefont
  {Bauer}},\ }\href {\doibase 10.1103/PhysRevLett.88.117601} {\bibfield
  {journal} {\bibinfo  {journal} {Phys. Rev. Lett.}\ }\textbf {\bibinfo
  {volume} {88}},\ \bibinfo {pages} {117601} (\bibinfo {year}
  {2002})}\BibitemShut {NoStop}%
\bibitem [{\citenamefont {Mosendz}\ \emph {et~al.}(2010)\citenamefont
  {Mosendz}, \citenamefont {Pearson}, \citenamefont {Fradin}, \citenamefont
  {Bauer}, \citenamefont {Bader},\ and\ \citenamefont
  {Hoffmann}}]{spin-pumping:Mosendz:PRL:2010}%
  \BibitemOpen
  \bibfield  {author} {\bibinfo {author} {\bibfnamefont {O.}~\bibnamefont
  {Mosendz}}, \bibinfo {author} {\bibfnamefont {J.~E.}\ \bibnamefont
  {Pearson}}, \bibinfo {author} {\bibfnamefont {F.~Y.}\ \bibnamefont {Fradin}},
  \bibinfo {author} {\bibfnamefont {G.~E.~W.}\ \bibnamefont {Bauer}}, \bibinfo
  {author} {\bibfnamefont {S.~D.}\ \bibnamefont {Bader}}, \ and\ \bibinfo
  {author} {\bibfnamefont {A.}~\bibnamefont {Hoffmann}},\ }\href {\doibase
  10.1103/PhysRevLett.104.046601} {\bibfield  {journal} {\bibinfo  {journal}
  {Phys. Rev. Lett.}\ }\textbf {\bibinfo {volume} {104}},\ \bibinfo {pages}
  {046601} (\bibinfo {year} {2010})}\BibitemShut {NoStop}%
\bibitem [{\citenamefont {Czeschka}\ \emph {et~al.}(2011)\citenamefont
  {Czeschka}, \citenamefont {Dreher}, \citenamefont {Brandt}, \citenamefont
  {Weiler}, \citenamefont {Althammer}, \citenamefont {Imort}, \citenamefont
  {Reiss}, \citenamefont {Thomas}, \citenamefont {Schoch}, \citenamefont
  {Limmer}, \citenamefont {Huebl}, \citenamefont {Gross},\ and\ \citenamefont
  {Goennenwein}}]{spin-pumping:Czeschka:scaling:PRL:2011}%
  \BibitemOpen
  \bibfield  {author} {\bibinfo {author} {\bibfnamefont {F.~D.}\ \bibnamefont
  {Czeschka}}, \bibinfo {author} {\bibfnamefont {L.}~\bibnamefont {Dreher}},
  \bibinfo {author} {\bibfnamefont {M.~S.}\ \bibnamefont {Brandt}}, \bibinfo
  {author} {\bibfnamefont {M.}~\bibnamefont {Weiler}}, \bibinfo {author}
  {\bibfnamefont {M.}~\bibnamefont {Althammer}}, \bibinfo {author}
  {\bibfnamefont {I.-M.}\ \bibnamefont {Imort}}, \bibinfo {author}
  {\bibfnamefont {G.}~\bibnamefont {Reiss}}, \bibinfo {author} {\bibfnamefont
  {A.}~\bibnamefont {Thomas}}, \bibinfo {author} {\bibfnamefont
  {W.}~\bibnamefont {Schoch}}, \bibinfo {author} {\bibfnamefont
  {W.}~\bibnamefont {Limmer}}, \bibinfo {author} {\bibfnamefont
  {H.}~\bibnamefont {Huebl}}, \bibinfo {author} {\bibfnamefont
  {R.}~\bibnamefont {Gross}}, \ and\ \bibinfo {author} {\bibfnamefont
  {S.~T.~B.}\ \bibnamefont {Goennenwein}},\ }\href {\doibase
  10.1103/PhysRevLett.107.046601} {\bibfield  {journal} {\bibinfo  {journal}
  {Phys. Rev. Lett.}\ }\textbf {\bibinfo {volume} {107}},\ \bibinfo {pages}
  {046601} (\bibinfo {year} {2011})}\BibitemShut {NoStop}%
\bibitem [{\citenamefont {Takei}\ \emph {et~al.}(2015)\citenamefont {Takei},
  \citenamefont {Moriyama}, \citenamefont {Ono},\ and\ \citenamefont
  {Tserkovnyak}}]{Takei:AFM:spincurrent:PRB:2015}%
  \BibitemOpen
  \bibfield  {author} {\bibinfo {author} {\bibfnamefont {S.}~\bibnamefont
  {Takei}}, \bibinfo {author} {\bibfnamefont {T.}~\bibnamefont {Moriyama}},
  \bibinfo {author} {\bibfnamefont {T.}~\bibnamefont {Ono}}, \ and\ \bibinfo
  {author} {\bibfnamefont {Y.}~\bibnamefont {Tserkovnyak}},\ }\href {\doibase
  10.1103/PhysRevB.92.020409} {\bibfield  {journal} {\bibinfo  {journal} {Phys.
  Rev. B}\ }\textbf {\bibinfo {volume} {92}},\ \bibinfo {pages} {020409}
  (\bibinfo {year} {2015})}\BibitemShut {NoStop}%
\bibitem [{\citenamefont {ichi Uchida}\ \emph {et~al.}(2010)\citenamefont {ichi
  Uchida}, \citenamefont {Adachi}, \citenamefont {Ota}, \citenamefont
  {Nakayama}, \citenamefont {Maekawa},\ and\ \citenamefont
  {Saitoh}}]{longitudinal-spin-Seebeck:Uchida:APL:2010:172505}%
  \BibitemOpen
  \bibfield  {author} {\bibinfo {author} {\bibfnamefont {K.}~\bibnamefont {ichi
  Uchida}}, \bibinfo {author} {\bibfnamefont {H.}~\bibnamefont {Adachi}},
  \bibinfo {author} {\bibfnamefont {T.}~\bibnamefont {Ota}}, \bibinfo {author}
  {\bibfnamefont {H.}~\bibnamefont {Nakayama}}, \bibinfo {author}
  {\bibfnamefont {S.}~\bibnamefont {Maekawa}}, \ and\ \bibinfo {author}
  {\bibfnamefont {E.}~\bibnamefont {Saitoh}},\ }\href {\doibase
  10.1063/1.3507386} {\bibfield  {journal} {\bibinfo  {journal} {Appl. Phys.
  Lett.}\ }\textbf {\bibinfo {volume} {97}},\ \bibinfo {eid} {172505} (\bibinfo
  {year} {2010})}\BibitemShut {NoStop}%
\bibitem [{\citenamefont {Weiler}\ \emph {et~al.}(2012)\citenamefont {Weiler},
  \citenamefont {Althammer}, \citenamefont {Czeschka}, \citenamefont {Huebl},
  \citenamefont {Wagner}, \citenamefont {Opel}, \citenamefont {Imort},
  \citenamefont {Reiss}, \citenamefont {Thomas}, \citenamefont {Gross},\ and\
  \citenamefont {Goennenwein}}]{Weiler:local:SSE:PRL:2012}%
  \BibitemOpen
  \bibfield  {author} {\bibinfo {author} {\bibfnamefont {M.}~\bibnamefont
  {Weiler}}, \bibinfo {author} {\bibfnamefont {M.}~\bibnamefont {Althammer}},
  \bibinfo {author} {\bibfnamefont {F.~D.}\ \bibnamefont {Czeschka}}, \bibinfo
  {author} {\bibfnamefont {H.}~\bibnamefont {Huebl}}, \bibinfo {author}
  {\bibfnamefont {M.~S.}\ \bibnamefont {Wagner}}, \bibinfo {author}
  {\bibfnamefont {M.}~\bibnamefont {Opel}}, \bibinfo {author} {\bibfnamefont
  {I.-M.}\ \bibnamefont {Imort}}, \bibinfo {author} {\bibfnamefont
  {G.}~\bibnamefont {Reiss}}, \bibinfo {author} {\bibfnamefont
  {A.}~\bibnamefont {Thomas}}, \bibinfo {author} {\bibfnamefont
  {R.}~\bibnamefont {Gross}}, \ and\ \bibinfo {author} {\bibfnamefont
  {S.~T.~B.}\ \bibnamefont {Goennenwein}},\ }\href {\doibase
  10.1103/PhysRevLett.108.106602} {\bibfield  {journal} {\bibinfo  {journal}
  {Phys. Rev. Lett.}\ }\textbf {\bibinfo {volume} {108}},\ \bibinfo {pages}
  {106602} (\bibinfo {year} {2012})}\BibitemShut {NoStop}%
\bibitem [{\citenamefont {Gepr\"{a}gs}\ \emph {et~al.}(2016)\citenamefont
  {Gepr\"{a}gs}, \citenamefont {Kehlberger}, \citenamefont {Coletta},
  \citenamefont {Qiu}, \citenamefont {Guo}, \citenamefont {Schulz},
  \citenamefont {Mix}, \citenamefont {Meyer}, \citenamefont {Kamra},
  \citenamefont {Althammer}, \citenamefont {Huebl}, \citenamefont {Jakob},
  \citenamefont {Ohnuma}, \citenamefont {Adachi}, \citenamefont {Barker},
  \citenamefont {Maekawa}, \citenamefont {Bauer}, \citenamefont {Saitoh},
  \citenamefont {Gross}, \citenamefont {Goennenwein},\ and\ \citenamefont
  {Klaui}}]{SSE:GdIG:Gepraegs:NatureCom:2016}%
  \BibitemOpen
  \bibfield  {author} {\bibinfo {author} {\bibfnamefont {S.}~\bibnamefont
  {Gepr\"{a}gs}}, \bibinfo {author} {\bibfnamefont {A.}~\bibnamefont
  {Kehlberger}}, \bibinfo {author} {\bibfnamefont {F.~D.}\ \bibnamefont
  {Coletta}}, \bibinfo {author} {\bibfnamefont {Z.}~\bibnamefont {Qiu}},
  \bibinfo {author} {\bibfnamefont {E.-J.}\ \bibnamefont {Guo}}, \bibinfo
  {author} {\bibfnamefont {T.}~\bibnamefont {Schulz}}, \bibinfo {author}
  {\bibfnamefont {C.}~\bibnamefont {Mix}}, \bibinfo {author} {\bibfnamefont
  {S.}~\bibnamefont {Meyer}}, \bibinfo {author} {\bibfnamefont
  {A.}~\bibnamefont {Kamra}}, \bibinfo {author} {\bibfnamefont
  {M.}~\bibnamefont {Althammer}}, \bibinfo {author} {\bibfnamefont
  {H.}~\bibnamefont {Huebl}}, \bibinfo {author} {\bibfnamefont
  {G.}~\bibnamefont {Jakob}}, \bibinfo {author} {\bibfnamefont
  {Y.}~\bibnamefont {Ohnuma}}, \bibinfo {author} {\bibfnamefont
  {H.}~\bibnamefont {Adachi}}, \bibinfo {author} {\bibfnamefont
  {J.}~\bibnamefont {Barker}}, \bibinfo {author} {\bibfnamefont
  {S.}~\bibnamefont {Maekawa}}, \bibinfo {author} {\bibfnamefont {G.~E.~W.}\
  \bibnamefont {Bauer}}, \bibinfo {author} {\bibfnamefont {E.}~\bibnamefont
  {Saitoh}}, \bibinfo {author} {\bibfnamefont {R.}~\bibnamefont {Gross}},
  \bibinfo {author} {\bibfnamefont {S.~T.~B.}\ \bibnamefont {Goennenwein}}, \
  and\ \bibinfo {author} {\bibfnamefont {M.}~\bibnamefont {Klaui}},\ }\href
  {\doibase 10.1038/ncomms10452} {\bibfield  {journal} {\bibinfo  {journal}
  {Nat. Commun.}\ }\textbf {\bibinfo {volume} {7}},\ \bibinfo {pages} {10452}
  (\bibinfo {year} {2016})}\BibitemShut {NoStop}%
\bibitem [{\citenamefont {Nakayama}\ \emph {et~al.}(2013)\citenamefont
  {Nakayama}, \citenamefont {Althammer}, \citenamefont {Chen}, \citenamefont
  {Uchida}, \citenamefont {Kajiwara}, \citenamefont {Kikuchi}, \citenamefont
  {Ohtani}, \citenamefont {Gepr\"ags}, \citenamefont {Opel}, \citenamefont
  {Takahashi}, \citenamefont {Gross}, \citenamefont {Bauer}, \citenamefont
  {Goennenwein},\ and\ \citenamefont {Saitoh}}]{NakayamaSMR}%
  \BibitemOpen
  \bibfield  {author} {\bibinfo {author} {\bibfnamefont {H.}~\bibnamefont
  {Nakayama}}, \bibinfo {author} {\bibfnamefont {M.}~\bibnamefont {Althammer}},
  \bibinfo {author} {\bibfnamefont {Y.-T.}\ \bibnamefont {Chen}}, \bibinfo
  {author} {\bibfnamefont {K.}~\bibnamefont {Uchida}}, \bibinfo {author}
  {\bibfnamefont {Y.}~\bibnamefont {Kajiwara}}, \bibinfo {author}
  {\bibfnamefont {D.}~\bibnamefont {Kikuchi}}, \bibinfo {author} {\bibfnamefont
  {T.}~\bibnamefont {Ohtani}}, \bibinfo {author} {\bibfnamefont
  {S.}~\bibnamefont {Gepr\"ags}}, \bibinfo {author} {\bibfnamefont
  {M.}~\bibnamefont {Opel}}, \bibinfo {author} {\bibfnamefont {S.}~\bibnamefont
  {Takahashi}}, \bibinfo {author} {\bibfnamefont {R.}~\bibnamefont {Gross}},
  \bibinfo {author} {\bibfnamefont {G.~E.~W.}\ \bibnamefont {Bauer}}, \bibinfo
  {author} {\bibfnamefont {S.~T.~B.}\ \bibnamefont {Goennenwein}}, \ and\
  \bibinfo {author} {\bibfnamefont {E.}~\bibnamefont {Saitoh}},\ }\href
  {\doibase 10.1103/PhysRevLett.110.206601} {\bibfield  {journal} {\bibinfo
  {journal} {Phys. Rev. Lett.}\ }\textbf {\bibinfo {volume} {110}},\ \bibinfo
  {pages} {206601} (\bibinfo {year} {2013})}\BibitemShut {NoStop}%
\bibitem [{\citenamefont {Althammer}\ \emph {et~al.}(2013)\citenamefont
  {Althammer}, \citenamefont {Meyer}, \citenamefont {Nakayama}, \citenamefont
  {Schreier}, \citenamefont {Altmannshofer}, \citenamefont {Weiler},
  \citenamefont {Huebl}, \citenamefont {Gepr\"ags}, \citenamefont {Opel},
  \citenamefont {Gross}, \citenamefont {Meier}, \citenamefont {Klewe},
  \citenamefont {Kuschel}, \citenamefont {Schmalhorst}, \citenamefont {Reiss},
  \citenamefont {Shen}, \citenamefont {Gupta}, \citenamefont {Chen},
  \citenamefont {Bauer}, \citenamefont {Saitoh},\ and\ \citenamefont
  {Goennenwein}}]{AltiPRB}%
  \BibitemOpen
  \bibfield  {author} {\bibinfo {author} {\bibfnamefont {M.}~\bibnamefont
  {Althammer}}, \bibinfo {author} {\bibfnamefont {S.}~\bibnamefont {Meyer}},
  \bibinfo {author} {\bibfnamefont {H.}~\bibnamefont {Nakayama}}, \bibinfo
  {author} {\bibfnamefont {M.}~\bibnamefont {Schreier}}, \bibinfo {author}
  {\bibfnamefont {S.}~\bibnamefont {Altmannshofer}}, \bibinfo {author}
  {\bibfnamefont {M.}~\bibnamefont {Weiler}}, \bibinfo {author} {\bibfnamefont
  {H.}~\bibnamefont {Huebl}}, \bibinfo {author} {\bibfnamefont
  {S.}~\bibnamefont {Gepr\"ags}}, \bibinfo {author} {\bibfnamefont
  {M.}~\bibnamefont {Opel}}, \bibinfo {author} {\bibfnamefont {R.}~\bibnamefont
  {Gross}}, \bibinfo {author} {\bibfnamefont {D.}~\bibnamefont {Meier}},
  \bibinfo {author} {\bibfnamefont {C.}~\bibnamefont {Klewe}}, \bibinfo
  {author} {\bibfnamefont {T.}~\bibnamefont {Kuschel}}, \bibinfo {author}
  {\bibfnamefont {J.-M.}\ \bibnamefont {Schmalhorst}}, \bibinfo {author}
  {\bibfnamefont {G.}~\bibnamefont {Reiss}}, \bibinfo {author} {\bibfnamefont
  {L.}~\bibnamefont {Shen}}, \bibinfo {author} {\bibfnamefont {A.}~\bibnamefont
  {Gupta}}, \bibinfo {author} {\bibfnamefont {Y.-T.}\ \bibnamefont {Chen}},
  \bibinfo {author} {\bibfnamefont {G.~E.~W.}\ \bibnamefont {Bauer}}, \bibinfo
  {author} {\bibfnamefont {E.}~\bibnamefont {Saitoh}}, \ and\ \bibinfo {author}
  {\bibfnamefont {S.~T.~B.}\ \bibnamefont {Goennenwein}},\ }\href {\doibase
  10.1103/PhysRevB.87.224401} {\bibfield  {journal} {\bibinfo  {journal} {Phys.
  Rev. B}\ }\textbf {\bibinfo {volume} {87}},\ \bibinfo {pages} {224401}
  (\bibinfo {year} {2013})}\BibitemShut {NoStop}%
\bibitem [{\citenamefont {Chen}\ \emph {et~al.}(2013)\citenamefont {Chen},
  \citenamefont {Takahashi}, \citenamefont {Nakayama}, \citenamefont
  {Althammer}, \citenamefont {Goennenwein}, \citenamefont {Saitoh},\ and\
  \citenamefont {Bauer}}]{ChenSMR}%
  \BibitemOpen
  \bibfield  {author} {\bibinfo {author} {\bibfnamefont {Y.-T.}\ \bibnamefont
  {Chen}}, \bibinfo {author} {\bibfnamefont {S.}~\bibnamefont {Takahashi}},
  \bibinfo {author} {\bibfnamefont {H.}~\bibnamefont {Nakayama}}, \bibinfo
  {author} {\bibfnamefont {M.}~\bibnamefont {Althammer}}, \bibinfo {author}
  {\bibfnamefont {S.~T.~B.}\ \bibnamefont {Goennenwein}}, \bibinfo {author}
  {\bibfnamefont {E.}~\bibnamefont {Saitoh}}, \ and\ \bibinfo {author}
  {\bibfnamefont {G.~E.~W.}\ \bibnamefont {Bauer}},\ }\href {\doibase
  10.1103/PhysRevB.87.144411} {\bibfield  {journal} {\bibinfo  {journal} {Phys.
  Rev. B}\ }\textbf {\bibinfo {volume} {87}},\ \bibinfo {pages} {144411}
  (\bibinfo {year} {2013})}\BibitemShut {NoStop}%
\bibitem [{\citenamefont {Hahn}\ \emph {et~al.}(2013)\citenamefont {Hahn},
  \citenamefont {de~Loubens}, \citenamefont {Klein}, \citenamefont {Viret},
  \citenamefont {Naletov},\ and\ \citenamefont
  {Ben~Youssef}}]{SMR:Hahn:PRB:2013}%
  \BibitemOpen
  \bibfield  {author} {\bibinfo {author} {\bibfnamefont {C.}~\bibnamefont
  {Hahn}}, \bibinfo {author} {\bibfnamefont {G.}~\bibnamefont {de~Loubens}},
  \bibinfo {author} {\bibfnamefont {O.}~\bibnamefont {Klein}}, \bibinfo
  {author} {\bibfnamefont {M.}~\bibnamefont {Viret}}, \bibinfo {author}
  {\bibfnamefont {V.~V.}\ \bibnamefont {Naletov}}, \ and\ \bibinfo {author}
  {\bibfnamefont {J.}~\bibnamefont {Ben~Youssef}},\ }\href {\doibase
  10.1103/PhysRevB.87.174417} {\bibfield  {journal} {\bibinfo  {journal} {Phys.
  Rev. B}\ }\textbf {\bibinfo {volume} {87}},\ \bibinfo {pages} {174417}
  (\bibinfo {year} {2013})}\BibitemShut {NoStop}%
\bibitem [{\citenamefont {Vlietstra}\ \emph {et~al.}(2013)\citenamefont
  {Vlietstra}, \citenamefont {Shan}, \citenamefont {Castel}, \citenamefont {van
  Wees},\ and\ \citenamefont {Ben~Youssef}}]{SMR:Vlietstra:PRB:2013}%
  \BibitemOpen
  \bibfield  {author} {\bibinfo {author} {\bibfnamefont {N.}~\bibnamefont
  {Vlietstra}}, \bibinfo {author} {\bibfnamefont {J.}~\bibnamefont {Shan}},
  \bibinfo {author} {\bibfnamefont {V.}~\bibnamefont {Castel}}, \bibinfo
  {author} {\bibfnamefont {B.~J.}\ \bibnamefont {van Wees}}, \ and\ \bibinfo
  {author} {\bibfnamefont {J.}~\bibnamefont {Ben~Youssef}},\ }\href {\doibase
  10.1103/PhysRevB.87.184421} {\bibfield  {journal} {\bibinfo  {journal} {Phys.
  Rev. B}\ }\textbf {\bibinfo {volume} {87}},\ \bibinfo {pages} {184421}
  (\bibinfo {year} {2013})}\BibitemShut {NoStop}%
\bibitem [{\citenamefont {Slonczewski}(1996)}]{Slonczewski:1996}%
  \BibitemOpen
  \bibfield  {author} {\bibinfo {author} {\bibfnamefont {J.~C.}\ \bibnamefont
  {Slonczewski}},\ }\href
  {http://www.sciencedirect.com/science/article/B6TJJ-403WKFH-1/2/fd06569c67005bde75d947ee6cfc273e}
  {\bibfield  {journal} {\bibinfo  {journal} {J. Magn. Magn. Mater.}\ }\textbf
  {\bibinfo {volume} {159}},\ \bibinfo {pages} {L1} (\bibinfo {year}
  {1996})}\BibitemShut {NoStop}%
\bibitem [{\citenamefont {Berger}(1996)}]{Berger1996}%
  \BibitemOpen
  \bibfield  {author} {\bibinfo {author} {\bibfnamefont {L.}~\bibnamefont
  {Berger}},\ }\href {\doibase 10.1103/PhysRevB.54.9353} {\bibfield  {journal}
  {\bibinfo  {journal} {Phys. Rev. B}\ }\textbf {\bibinfo {volume} {54}},\
  \bibinfo {pages} {9353} (\bibinfo {year} {1996})}\BibitemShut {NoStop}%
\bibitem [{\citenamefont {Chen}\ \emph {et~al.}(2016)\citenamefont {Chen},
  \citenamefont {Takahashi}, \citenamefont {Nakayama}, \citenamefont
  {Althammer}, \citenamefont {Goennenwein}, \citenamefont {Saitoh},\ and\
  \citenamefont {Bauer}}]{SMR:review:Chen:JPCM:2016}%
  \BibitemOpen
  \bibfield  {author} {\bibinfo {author} {\bibfnamefont {Y.-T.}\ \bibnamefont
  {Chen}}, \bibinfo {author} {\bibfnamefont {S.}~\bibnamefont {Takahashi}},
  \bibinfo {author} {\bibfnamefont {H.}~\bibnamefont {Nakayama}}, \bibinfo
  {author} {\bibfnamefont {M.}~\bibnamefont {Althammer}}, \bibinfo {author}
  {\bibfnamefont {S.~T.~B.}\ \bibnamefont {Goennenwein}}, \bibinfo {author}
  {\bibfnamefont {E.}~\bibnamefont {Saitoh}}, \ and\ \bibinfo {author}
  {\bibfnamefont {G.~E.~W.}\ \bibnamefont {Bauer}},\ }\href {\doibase
  10.1038/ncomms10452} {\bibfield  {journal} {\bibinfo  {journal} {J. Phys.:
  Condens. Matter}\ }\textbf {\bibinfo {volume} {28}},\ \bibinfo {pages}
  {103004} (\bibinfo {year} {2016})}\BibitemShut {NoStop}%
\bibitem [{\citenamefont {Jia}\ \emph {et~al.}(2011)\citenamefont {Jia},
  \citenamefont {Liu}, \citenamefont {Xia},\ and\ \citenamefont
  {Bauer}}]{Jia:2011gm}%
  \BibitemOpen
  \bibfield  {author} {\bibinfo {author} {\bibfnamefont {X.}~\bibnamefont
  {Jia}}, \bibinfo {author} {\bibfnamefont {K.}~\bibnamefont {Liu}}, \bibinfo
  {author} {\bibfnamefont {K.}~\bibnamefont {Xia}}, \ and\ \bibinfo {author}
  {\bibfnamefont {G.~E.~W.}\ \bibnamefont {Bauer}},\ }\href@noop {} {\bibfield
  {journal} {\bibinfo  {journal} {Europhys. Lett.}\ }\textbf {\bibinfo {volume}
  {96}},\ \bibinfo {pages} {17005} (\bibinfo {year} {2011})}\BibitemShut
  {NoStop}%
\bibitem [{\citenamefont {Bernasconi}\ and\ \citenamefont
  {Kuse}(1971)}]{Bernasconi}%
  \BibitemOpen
  \bibfield  {author} {\bibinfo {author} {\bibfnamefont {J.}~\bibnamefont
  {Bernasconi}}\ and\ \bibinfo {author} {\bibfnamefont {D.}~\bibnamefont
  {Kuse}},\ }\href {\doibase 10.1103/PhysRevB.3.811} {\bibfield  {journal}
  {\bibinfo  {journal} {Phys. Rev. B}\ }\textbf {\bibinfo {volume} {3}},\
  \bibinfo {pages} {811} (\bibinfo {year} {1971})}\BibitemShut {NoStop}%
\bibitem [{\citenamefont {Clark}\ and\ \citenamefont {Callen}(1968)}]{Clark}%
  \BibitemOpen
  \bibfield  {author} {\bibinfo {author} {\bibfnamefont {A.~E.}\ \bibnamefont
  {Clark}}\ and\ \bibinfo {author} {\bibfnamefont {E.}~\bibnamefont {Callen}},\
  }\href {\doibase 10.1063/1.1656100} {\bibfield  {journal} {\bibinfo
  {journal} {Jour. Appl. Phys.}\ }\textbf {\bibinfo {volume} {39}},\ \bibinfo
  {pages} {5972} (\bibinfo {year} {1968})}\BibitemShut {NoStop}%
\bibitem [{\citenamefont {Miura}\ \emph {et~al.}(1978)\citenamefont {Miura},
  \citenamefont {Oguro},\ and\ \citenamefont {Chikazumi}}]{Chikazumi}%
  \BibitemOpen
  \bibfield  {author} {\bibinfo {author} {\bibfnamefont {N.}~\bibnamefont
  {Miura}}, \bibinfo {author} {\bibfnamefont {I.}~\bibnamefont {Oguro}}, \ and\
  \bibinfo {author} {\bibfnamefont {S.}~\bibnamefont {Chikazumi}},\ }\href
  {\doibase 10.1143/JPSJ.45.1534} {\bibfield  {journal} {\bibinfo  {journal}
  {Jour. Phys. Soc. Jap.}\ }\textbf {\bibinfo {volume} {45}},\ \bibinfo {pages}
  {1534} (\bibinfo {year} {1978})}\BibitemShut {NoStop}%
\bibitem [{\citenamefont {Hinzke}\ and\ \citenamefont
  {Nowak}(1999)}]{Hinzke:1999ud}%
  \BibitemOpen
  \bibfield  {author} {\bibinfo {author} {\bibfnamefont {D.}~\bibnamefont
  {Hinzke}}\ and\ \bibinfo {author} {\bibfnamefont {U.}~\bibnamefont {Nowak}},\
  }\href@noop {} {\bibfield  {journal} {\bibinfo  {journal} {Comput. Phys.
  Commun.}\ }\textbf {\bibinfo {volume} {121}},\ \bibinfo {pages} {334}
  (\bibinfo {year} {1999})}\BibitemShut {NoStop}%
\bibitem [{\citenamefont {Meyer}\ \emph {et~al.}(2014)\citenamefont {Meyer},
  \citenamefont {Althammer}, \citenamefont {Gepr{\"a}gs}, \citenamefont {Opel},
  \citenamefont {Gross},\ and\ \citenamefont {Goennenwein}}]{MeyerSMRAPL}%
  \BibitemOpen
  \bibfield  {author} {\bibinfo {author} {\bibfnamefont {S.}~\bibnamefont
  {Meyer}}, \bibinfo {author} {\bibfnamefont {M.}~\bibnamefont {Althammer}},
  \bibinfo {author} {\bibfnamefont {S.}~\bibnamefont {Gepr{\"a}gs}}, \bibinfo
  {author} {\bibfnamefont {M.}~\bibnamefont {Opel}}, \bibinfo {author}
  {\bibfnamefont {R.}~\bibnamefont {Gross}}, \ and\ \bibinfo {author}
  {\bibfnamefont {S.~T.~B.}\ \bibnamefont {Goennenwein}},\ }\href {\doibase
  10.1063/1.4885086} {\bibfield  {journal} {\bibinfo  {journal} {Appl. Phys.
  Lett.}\ }\textbf {\bibinfo {volume} {104}},\ \bibinfo {eid} {242411}
  (\bibinfo {year} {2014})}\BibitemShut {NoStop}%
\bibitem [{\citenamefont {Aqeel}\ \emph {et~al.}(2015)\citenamefont {Aqeel},
  \citenamefont {Vlietstra}, \citenamefont {Heuver}, \citenamefont {Bauer},
  \citenamefont {Noheda}, \citenamefont {van Wees},\ and\ \citenamefont
  {Palstra}}]{SMR:CCO:Aqeel:PRB2015}%
  \BibitemOpen
  \bibfield  {author} {\bibinfo {author} {\bibfnamefont {A.}~\bibnamefont
  {Aqeel}}, \bibinfo {author} {\bibfnamefont {N.}~\bibnamefont {Vlietstra}},
  \bibinfo {author} {\bibfnamefont {J.~A.}\ \bibnamefont {Heuver}}, \bibinfo
  {author} {\bibfnamefont {G.~E.~W.}\ \bibnamefont {Bauer}}, \bibinfo {author}
  {\bibfnamefont {B.}~\bibnamefont {Noheda}}, \bibinfo {author} {\bibfnamefont
  {B.~J.}\ \bibnamefont {van Wees}}, \ and\ \bibinfo {author} {\bibfnamefont
  {T.~T.~M.}\ \bibnamefont {Palstra}},\ }\href {\doibase
  10.1103/PhysRevB.92.224410} {\bibfield  {journal} {\bibinfo  {journal} {Phys.
  Rev. B}\ }\textbf {\bibinfo {volume} {92}},\ \bibinfo {pages} {224410}
  (\bibinfo {year} {2015})}\BibitemShut {NoStop}%
\end{thebibliography}
%

\end{document}